# COVID-19 and India: What Next?

Ramesh Behl[1] and Manit Mishra[2]

**Abstract**

The study carries out predictive modeling based on publicly available COVID-19 data for the duration 01 April to 20 June 2020 pertaining to India and five of its most infected states: Maharashtra, Tamil Nadu, Delhi, Gujarat, and Rajasthan using susceptible, infected, recovered, and dead (SIRD) model. The basic reproduction number $R_0$ is derived by exponential growth method using RStudio package R0. The differential equations reflecting SIRD model have been solved using Python 3.7.4 on Jupyter Notebook platform. For visualization, Python Matplotlib 3.2.1 package is used. The study offers insights on peak-date, peak number of COVID-19 infections, and end-date pertaining to India and five of its states. The results could be leveraged by political leadership, health authorities, and industry doyens for policy planning and execution.

(1) Professor and Director, International Management Institute Bhubaneswar, India, E-mail id: *rbehl@imi.edu*

(2) Associate Professor, International Management Institute Bhubaneswar, India, E-mail id: *manit.mishra@imibh.edu.in*

# 1. Introduction

Quantitative conjectures about spread of epidemics or pandemics based on mathematical modeling using simulations and data analysis techniques is critical information for policy makers as well as ordinary citizens susceptible to these viral infections. Even as the pandemic batters the whole world, India is facing unique challenges. India imposed country-wide lock down on 24th March 2020 which continued up to 17th May 2020. On 17 May 2020 lockdown was further extended up to 31st May 2020 with different rules for three different district zones – green, red, and orange – created based on spread of pandemic. The lockdown in India during 25 March to 31 May 2020 has been acknowledged as one of the most stringent in the world (Hale et al., 2020). Researchers have also recognized the fact that but for the lockdown, the number of infected individuals in India would have been much larger. A study done by The Center for Disease Dynamics, Economics & Policy (CEDEP), John Hopkins, and Princeton suggests that the very first phase of 21-day nationwide lockdown up to 14 April 2020 must have significantly slowed down the spread of COVID-19 as compared to a lack of intervention (Schuller et al., 2020). 1st of June 2020 onwards, the Government of India issued the notification for "Unlock 1" with emphasis on restarting economic activities and easing of restrictions in all areas except the containment zones. The Unlock 1 phase has started even as the cumulative number of confirmed COVID-19 cases and per day increase in confirmed cases in India (Figure 1) is still showing an upward trend. As on 20th June 2020 the number of confirmed cases in India stood at 4,11,727 (Worldometers, 2020, June 21) with the per day increase in number of cases reaching 14,516 (WHO, 2020, June 21).

      The five Indian states analyzed in this study also show an increasing trend in number of confirmed cases. As a result, India is caught in a catch-22 situation having to choose between the devil and the deep sea. It has come down to an argument between life and livelihood.

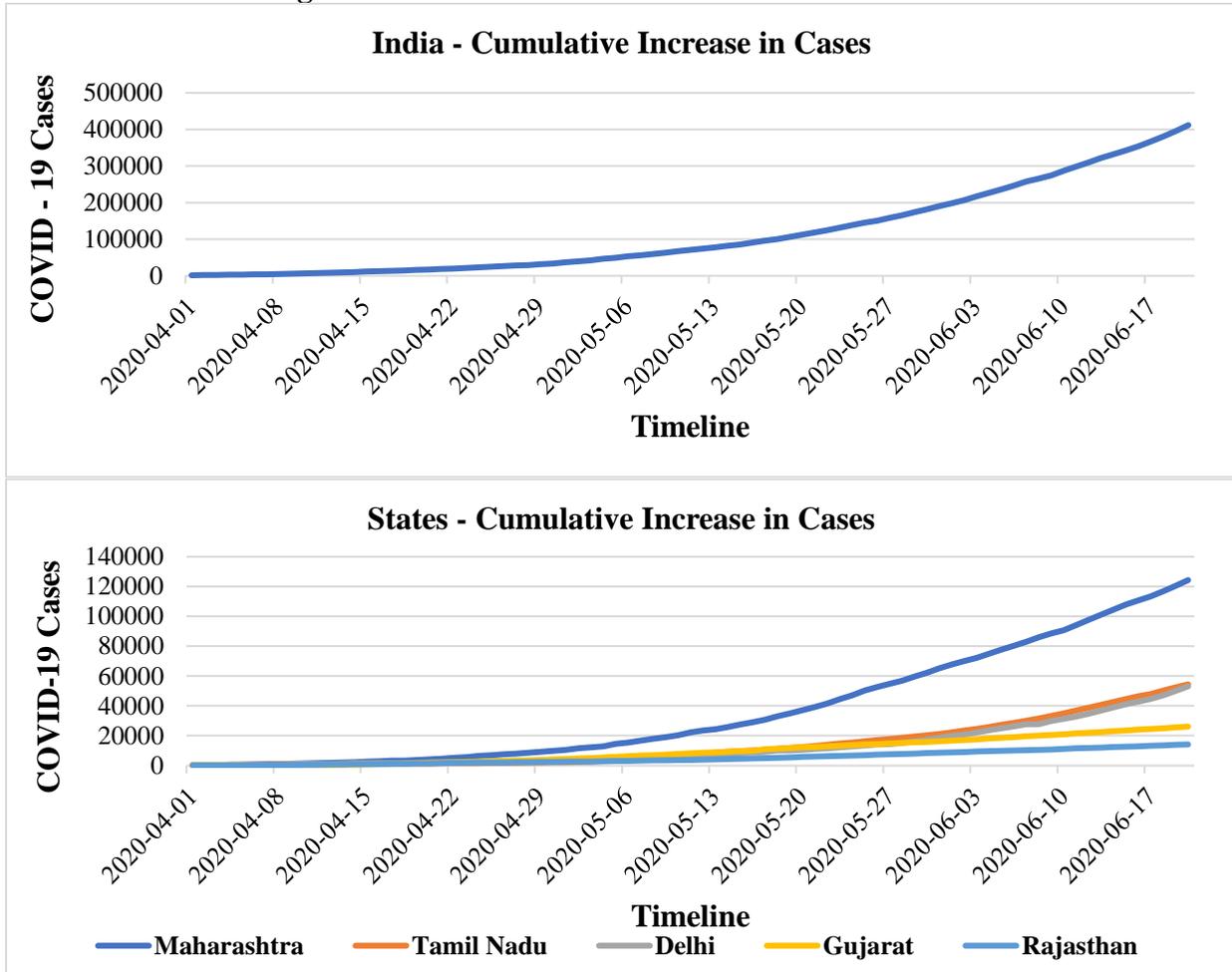

**Figure 1.** Trend of Cumulative Cases – India and States

While the number of confirmed cases is still registering an exponential growth, the lockdown is being eased to restart economic activities. In such a scenario, the study explores the longevity of COVID-19 in India and five of its states using confirmed cases data and based on a robust theoretical framework – the susceptible, infected, recovered, and dead (SIRD) compartmental model. Our analysis is of utility to policy makers, public health authorities, and industry leaders to prepare themselves better in the coming months in terms of healthcare resources and restarting of economic activities. It also enlightens the common citizens regarding COVID-19 timelines so that they are mentally prepared for a self-protective lifestyle. The study, thus, fulfills two objectives. First, we predict the COVID-19 longevity timelines for India using confirmed cases time series data during 01 April to 20 June 2020 based on SIRD model. Second, we also derive the COVID-19 longevity pattern for five most infected states which cumulatively contribute approximately 70% cases to the India tally. These states are Maharashtra, Tamil Nadu, Delhi, Gujarat, and Rajasthan. In the process, we predict the following attributes related to the COVID-19 lifespan pattern of India and its states: (1) the peak-date; (2) the peak number of infections; and, (3) the end-date.

To the best of our knowledge, no study so far has endeavored to fulfill these objectives so comprehensively. Even though Behl and Mishra (2020) have attempted to predict lifecycle patterns of COVID-19, our study is different from theirs on following four counts. First, our study's theoretical foundation is based on SIRD model whereas their study is based on susceptible, infected, removed (SIR) model. Second, we derive the predictions based on confirmed case data of more than two months (01 April to 20 June 2020) whereas they used the data of only one month (01 to 30 April 2020). Moreover, it was at a time when the trend was just about beginning to exhibit an upward trend. Third, our study predicts COVID-19 lifecycle pattern of India as a whole, along with five of its most affected states whereas their study carried out prediction for ten most infected states only. Fourth, unlike their study, we derive the critical parameter of basic reproduction number $R_0$ for India and each state separately using the exponential growth method proposed by Wallinga and Lipsitch (2007). Even as we make a quantitative conjecture in this article regarding the lifecycle of COVID-19, we must concede that we are not epidemiologists. We are data scientists reporting what the data is telling us. Therefore, the results should be interpreted and used with caution. The remaining part of the article comprises of following sections: Section 2 discusses the conceptual framework; Section 3 describes the research method; Section 4 presents outputs of the study, and finally; Section 5 delves upon on implications.

## 2. Conceptual Framework

Extant literature in Epidemiology has used various compartmental models to analyze the spread of infectious diseases (Hethcote, 2000). The most prominent and widely used among them being SIR model proposed by Kermack and McKendrick (1927). However, SIR model is criticized as being too simplistic (Bhattacharya et al., 2015) and is limited by the assumption that the pandemic is non-fatal (Miller, 2020). With the fatalities due to COVID-19 going up, it would not be suitable to assume zero fatalities. Therefore, this study draws its theoretical moorings from the seminal SIRD model of epidemiology for predictive modeling of COVID-19 in India. The SIRD model has been extensively used in extant literature to model spread of epidemics, in general, and COVID-19, in particular (e.g., Caccavo, 2020; Chatterjee et al., 2020; Fernández-Villaverde and Jones, 2020). The SIRD model classifies individuals into four mutually exclusive compartments: susceptible (S), infected (I), recovered (R), and dead (D). The SIRD model considers recovered and deceased as separate compartments and, therefore, must consider the case-fatality rate. The transition from susceptible to infected, and from infected to either recovered or dead is captured through a series of differential equations.

$$ds/dt = - \beta\,(si)$$

$$di/dt = \beta\,(si) - (\acute{v} + \sigma)\,i$$

$$dr/dt = \acute{v}\,(i)$$

$$dd/dt = \sigma\,(i)$$

Wherein, $s = S/N$, $i = I/N$, $r = R/N$, and $d = D/N$. Here, $\beta$ is the effective contact rate which captures the proportion of susceptible population in contact that an infected individual further infects per unit time (Hill, 2016). $\acute{v}$, the removal rate, is the inverse of the expected duration of infection (Jones, 2007). And $\sigma$ is the case-fatality rate i.e. the number of infected patients who die per unit of time (Miller, 2020). The other acronyms used in the equations are N, the population; S, the number of susceptible individuals at time *t*; I, the number of infected individuals at time *t*; R, the number of individuals who have recovered from infection at time *t*; and, D is the number of deceased individuals at time *t*. The unit of time considered is day and therefore, all calculations have been done on per day basis. Additionally, the population N need not necessarily be same as the population of the state (Batista, 2020). If we consider all members of this population equally likely to be infected, N can be considered as equal to the total susceptible population at the beginning of time. This study is based on reported confirmed cases of COVID-19 (C). Therefore, as C increases, the population of susceptible individuals S out of the total population N keeps decreasing and moving away from N. The data on number of confirmed cases C includes infected, recovered, as well as deceased individuals. The application of SIRD model is based on two important assumptions: (1) the model is recursive and therefore, there is no possibility of a recovered individual getting infected again; and, (2) the population is constant during the period of study which means the birth rate and death rate, including deaths due to COVID-19, are equal (Miller, 2020). The laws of motion with respect to COVID-19 are as depicted in Figure 2.

**Figure 2.** Schematic Representation of SIRD Model

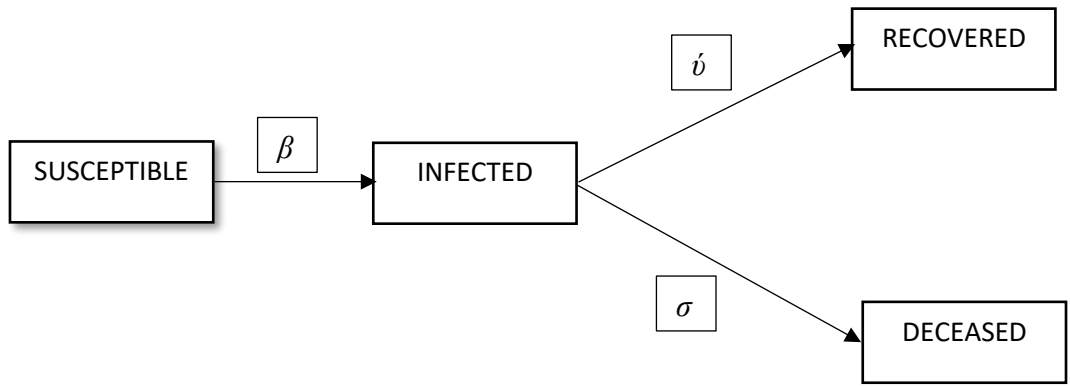

## 3. Research Design

### *3.1 Data*

We have taken into consideration COVID-19 confirmed cases data for the duration 01 April to 20 June 2020. The time series data reports per-day wise change in the number of cases. The data pertains to India and five of its most infected states which together contribute approximately 70% of the cases as on 20 June 2020 – Maharashtra, Tamil Nadu, Delhi, Gujarat, and Rajasthan. The state-wise time-series data was downloaded from Kaggle (Rajkumar, 2020) whereas the time-series data pertaining to India was downloaded from Our World in Data (Roser et al., 2020). Both sources are publicly accessible data platforms. Even though the first case in India was reported on 31 January 2020, the data prior to $1^{st}$ April 2020 was not utilized to avoid getting trapped by the noise in the data.

### *3.2 Methodology*

The critical parameters used for predictive modeling are determined as per established procedures (Batista, 2020; Behl and Mishra, 2020).

#### *3.2.1 Susceptible Population (S)*

The total population, denoted by N, is also equal to the population of individuals susceptible (S) to infection at the beginning of the time period. The value of susceptible population is estimated using the method proposed by Chatterjee et al. (2020). We have estimated the lower bound and upper bound of susceptible population and thereafter taken an average of it. The lower bound of susceptible population is the number of individuals infected as on 20 June 2020. On the other hand, the upper bound has been estimated based on positivity ratio, which is derived from the ratio between the number of infected individuals and the number of tests done. The positivity ratio is multiplied to the actual population of the country/state to estimate the upper bound of population size vulnerable to infection. The susceptible population size is estimated as the average of the upper and lower bounds (Table I).

**Table I.** Determining the Susceptible Population

| Country/State | Population (In Crores) | Number of tests[#] | Positivity Ratio[#] | Upper Bound of S | Lower bound of S (Number of positive cases) | Susceptible Population (S) |
|---|---|---|---|---|---|---|
| India | 130 | 66,16496 | 0.06 | 8,84,00,000 | 4,11,752 | 4,44,05,864 |
| Maharashtra | 11.4 | 7,56,809 | 0.17 | 1,93,80,000 | 1,28,205 | 97,54,102 |
| Tamil Nadu | 6.79 | 8,61,211 | 0.06 | 46,17,200 | 56,845 | 23,37,022 |
| Delhi | 1.9 | 3,51,909 | 0.16 | 30,40,000 | 56,746 | 15,48,373 |
| Gujarat | 6.27 | 3,19,414 | 0.08 | 50,16,000 | 26,737 | 25,21,369 |
| Rajasthan | 6.89 | 6,83,017 | 0.02 | 13,78,000 | 14,537 | 6,96,269 |

Source: The authors. # Data obtained from covid19india.org. Positivity ratio is the ratio of COVID-19 positive cases to number of tests done as on 20[th] June 2020.

*3.2.2 Basic Reproduction Number ($R_0$)*

An important epidemiological parameter in modeling of infectious diseases is the basic reproduction number denoted by $R_0$ (Bettencourt and Ribeiro, 2008). The basic reproduction number for a pandemic in each region is determined based on the "expected number of secondary cases produced by a single (typical) infection in a completely susceptible population" (Jones, 2007, p. 1). Liu et al. (2020) argue that an $R_0$ value greater than one suggests rapid transmission of the infectious disease whereas if the value of $R_0$ is less than one, it indicates a decreasing transmission. For COVID-19, the World Health Organization (WHO) suggests an $R_0$ ranging between 1.4 to 2.5 (Liu et al. 2020). Some of the other estimated range of values of $R_0$ for different countries are: 2.6 to 3.2 for South Korea and 2.6 to 3.3 for Italy (Zhuang et al., 2020); and, 2.24 to 3.58 for China (Zhao et al., 2020). Liu et al. (2020) did a meta-analysis of 12 studies which estimated $R_0$ for COVID-19 in various parts of the world and suggested the mean $R_0$ value of 3.28. For this study we estimate the value of $R_0$ by applying exponential growth method recommended by Wallinga and Lipsitch (2007). We used RStudio 1.3.959 (RStudio Inc., Boston, MA) package R0 1.2-6 (Obadia et al., 2012) to estimate the value of $R_0$ for India and five of its states. The parameter case generation time, which is required for calculation of $R_0$, is estimated from the mean serial interval value. The serial interval is the time gap between onset of illness in a primary and secondary case. For COVID-19, Bi et al. (2020) estimate the mean serial interval value to be 6.3 days with a standard deviation of 4.2 days.

*3.2.3 The recovery rate (ύ)*

Even as research on recovery period from COVID-19 is still in its infancy, most experts agree on a 2 to 6-week period. World Health Organization (WHO) suggests that 80-85% of COVID-19 patients recover in about 2-weeks. This contention to a large extent is also supported by academic research (Toda, 2020). For India, Chatterjee et al. (2020) suggest a recovery period of 12 days. Based on the preceding discussion, we have considered a recovery period of 12 days.

*3.2.4 The effective contact rate (β)*

The effective contact rate, denoted by *β*, influences the spread of the epidemic. It is defined as per capita "average number of effective contacts with other individual hosts per unit time" (Li et al., 1999, p. 192). Among the three parameters of the SIRD model, this is the only parameter that is controllable and therefore, can check the spread of epidemic (Toda, 2020). The effective contact rate of an epidemic in a population is determined by two factors – the average rate of contact between infected and susceptible individuals i.e. the total number of contacts per unit time, and the transmissibility of the epidemic (Jones, 2007). The effective contact rate, rate of recovery, and basic reproduction number of an epidemic are related to each other by the following equation, as suggested in extant research (Bhattacharya et al., 2015; Jones, 2007). This equation was used to calculate the values of effective contact rate $β$ for India and the five states modeled in this study.

$$R_0 = (β/\acute{v}) \quad\text{———} \quad (1)$$

*3.2.5 The case-fatality rate (σ)*

The case-fatality rate $σ$ reflects the number of infected patients who die per unit time i.e. each day. It is related to recovery rate $\acute{v}$ and fraction of infected individuals who die $R_1$ in the following equation (Miller, 2020). The value of $σ$ varies across the states considered and is different from the national value.

$$(σ = R_1*\acute{v}/100) \quad\text{———} \quad (2)$$

## 4. Results

The implementation of SIRD model depends on the parameters effective contact rate ($β$), rate of recovery ($\acute{v}$), and the case-fatality rate ($σ$). The effective contact rate, in turn, depends upon the basic reproduction number ($R_0$) of the epidemic (Jones, 2007). We estimated the values of basic reproduction number $R_0$ by applying exponential growth method (Wallinga and Lipsitch, 2007) using RStudio 1.3.959 (RStudio Inc., Boston, MA) package R0 1.2-6 (Obadia et al., 2012). The exponential growth rate, calculated by applying exponential function to the data, of COVID-19 confirmed cases in India was found to be 1.05, 1.04, and 1.03 for April, May, and June (01-20 June), respectively. Obadia et al. (2012, p. 2) argue that "It is necessary to choose a period in the epidemic curve over which growth is exponential." Therefore, we choose data from April to compute the value of the basic reproduction number. The $R_0$ values obtained for India, Maharashtra, Tamil Nadu, Delhi, Gujarat, and Rajasthan are 1.72, 1.90, 1.45, 1.62, 2.14, and 1.73, respectively. The values obtained for $R_0$ are within the WHO recommended range of 1.4 to 2.5 (Liu et al., 2020).

The recovery rate ($\acute{v}$) of COVID-19 infected individuals is determined by the average recovery period. The recovery rate is considered as inversely proportional to the recovery period in days (Hill, 2016; Toda, 2020). Since the recovery period of COVID-19 is 12 days (Chatterjee, 2020), a uniform recovery rate of $\acute{v} = 1/12 = 0.08$ is considered for India as well as all five of its states. The value of recovery rate is expected to be constant irrespective of the geographical region. The recovery rate obtained is used to calculate the values of effective contact rate $\beta$ using equation (1) and the case-fatality rate $\sigma$ using equation (2). The values of susceptible population (S), basic reproduction number ($R_0$), recovery rate ($\acute{v}$), effective contact rate ($\beta$), proportion of deaths among infected individuals ($R_1$), and case-fatality rate ($\sigma$) for India and the five states investigated are provided in Table II.

**Table II**. Determining the SIRD model parameters

| # | Country/ State | Susceptible Population (S)# | Basic reproduction number ($R_0$) [95% CI] | Recovery rate ($\acute{v}$) | Effective contact rate ($\beta$) | Percentage of deaths ($R_1$)* | Case-fatality rate ($\sigma$) |
|---|---|---|---|---|---|---|---|
| 1 | India | 44405864 | 1.719 [1.715, 1.723] | 0.08 | 0.14 | 3.22 | 0.0026 |
| 2 | Maharashtra | 9754102 | 1.900 [1.891, 1.909] | 0.08 | 0.15 | 4.74 | 0.0038 |
| 3 | Tamil Nadu | 2337022 | 1.446 [1.435, 1.457] | 0.08 | 0.12 | 1.22 | 0.0010 |
| 4 | Delhi | 1548373 | 1.622 [1.611, 1.633] | 0.08 | 0.13 | 3.83 | 0.0031 |
| 5 | Gujarat | 2521369 | 2.144 [2.127, 2.161] | 0.08 | 0.17 | 6.19 | 0.0050 |
| 6 | Rajasthan | 696269 | 1.731 [1.717, 1.745] | 0.08 | 0.14 | 2.35 | 0.0019 |

Source: The authors. *Calculated by authors based on data obtained from Kaggle (Rajkumar, 2020) and https://www.worldometers.info/coronavirus/. #Calculated from the population, as per the process given in Table I and therefore, no confidence intervals are provided.

**Figure 3.** COVID-19 SIRD Models

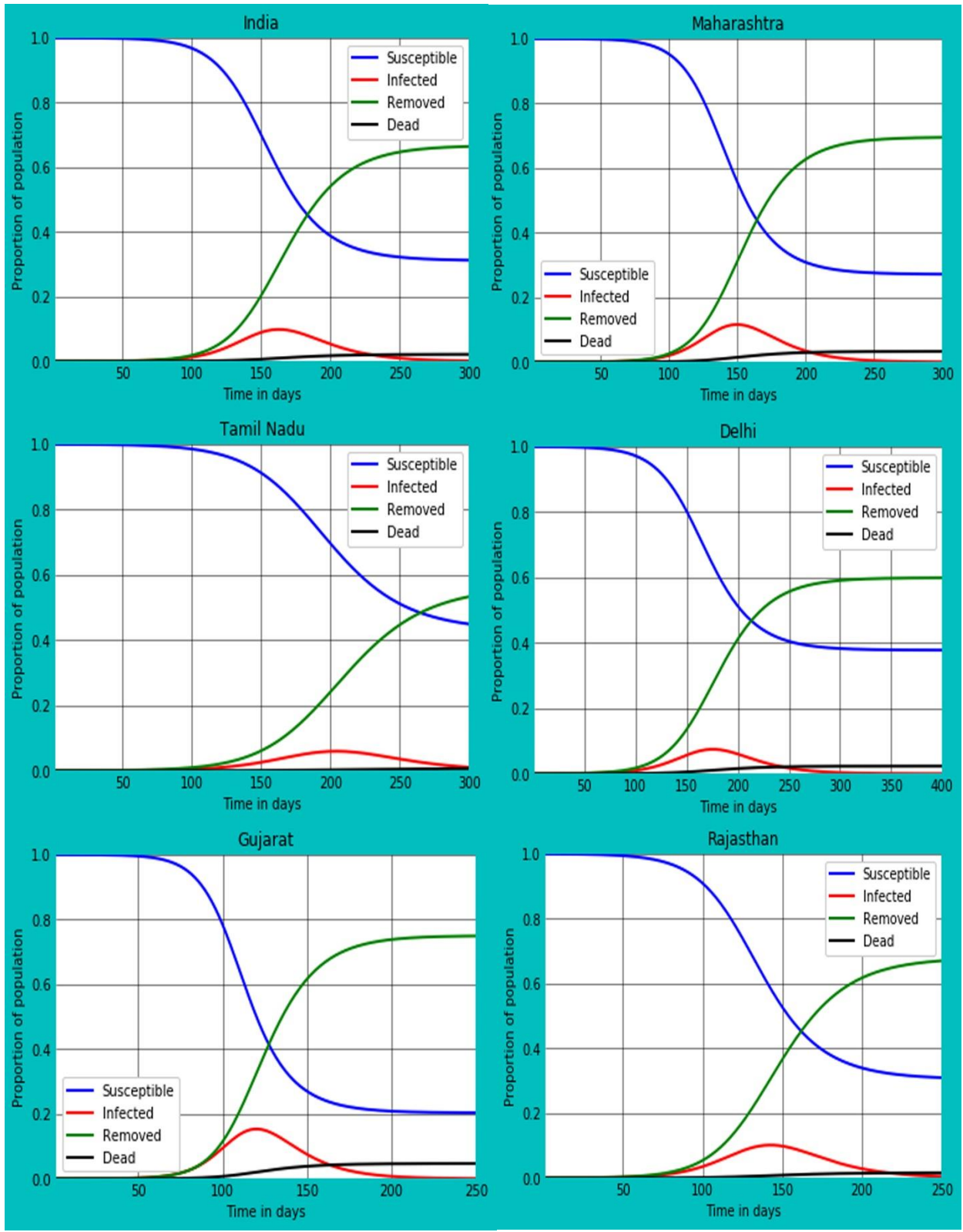

The values of initially susceptible population (S); number of infected (I), recovered (R), and dead (D) individuals; and, effective contact rate (*β*), rate of recovery (*ύ*), and the case-fatality rate (*σ*) are the input parameters for running SIRD model. The set of ordinary differential equations defining SIRD model are run on Python 3.7.4 (Van Rossum and Drake, 2009) using the Jupyter Notebook (Kluyver et al., 2016) platform. The COVID-19 timeline plots for India and the states of Maharashtra, Tamil Nadu, Delhi, Gujarat, Rajasthan are obtained using the Python package Matplotlib 3.2.1 (Hunter, 2007). The predictive modeling plots (Figure 3) provide insights pertaining to: (1) the peak-date and peak cumulative number of infected individuals in the susceptible population; and, (2) the end-date of pandemic in India and five of its states. The information in the plots pertaining to peak date, number of cumulative infections as on peak date and the end date have been tabulated (Table III) for better understanding.

**Table III**. Predicting peak date and end date

| # | Country/State | Peak date | Predicted number of confirmed infections as on peak date | 95% Confidence interval | | End date |
|---|---|---|---|---|---|---|
| | | | | Lower | Upper | |
| 1 | India | 160 days-10 September 20 | 44,40,586 | 32,17,435 | 56,63,737 | 300 days-31 January 21 |
| 2 | Maharashtra | 150 days-30 August 20 | 11,70,492 | 8,26,633 | 15,14,351 | 270 days-31 December 20 |
| 3 | Tamil Nadu | 200 days-20 October 20 | 1,40,221 | 1,11,161 | 1,69,281 | 360 days-30 March 20 |
| 4 | Delhi | 175 days-25 September 20 | 1,16,128 | 88,626 | 1,43,630 | 325 days-25 February 20 |
| 5 | Gujarat | 120 days – 31 July 20 | 3,78,205 | 2,60,998 | 4,95,412 | 225 days-15 November 20 |
| 6 | Rajasthan | 140 days – 20 August 20 | 69,627 | 51,021 | 88,233 | 275 days-05 January 20 |

## 5. Discussion and conclusion

Quantitative conjectures based on mathematical modeling play an important role in understanding an epidemic. Drawing upon SIRD model, this study endeavors to do predictive modeling of COVID-19 epidemic in India and five of its states: Maharashtra, Tamil Nadu, Delhi, Gujarat, and Rajasthan. These are the five most infected states and account for approximately 70% of all COVID-19 confirmed cases in India as on 20$^{th}$ June 2020 as per data obtained from Kaggle (Rajkumar, 2020). It is a scientifically robust study wherein all parameters have been estimated objectively rather than relying excessively on assumptions. We have used established tools and techniques to arrive at our estimates. Population susceptible (S) to COVID-19 has been determined by averaging the lower-bound and upper-bound estimates as proposed by Chatterjee et al. (2020). The basic reproduction number ($R_0$) has been calculated using RStudio 1.3.959 (RStudio Inc., Boston, MA) package R0 1.2-6 (Obadia et al., 2012) through the exponential growth method (Wallinga and Lipsitch, 2007). The recovery rate ($ύ$) has been calculated from the recovery period of COVID-19 available in extant literature (Chatterjee et al., 2020). The effective contact rate ($β$) and case-fatality rate ($σ$) have been obtained by using equations (1) and (2), respectively. Finally, the ordinary differential equations defining SIRD model are run on Python 3.7.4 (Van Rossum and Drake, 2009) using the Jupyter Notebook (Kluyver et al., 2016) platform and the plots (Figure 3) are obtained using Python Matplotlib 3.2.1 (Hunter, 2007) package.

      The projections from this study throw up two interesting insights for policy makers, industry leaders, and healthcare experts. The basic reproduction number $R_0$ for India was found to 1.72. Even as there is considerable debate on $R_0$, a relatively lower value of 1.72 obtained in this study for India in comparison to global estimates (3.28 and 3.32) is an encouraging sign. Liu et al. (2020) estimated the mean value of $R_0$ to be around 3.28 based on a meta-analysis of twelve studies

conducted during 31 December 2019 to 22 January 2020. Another meta-analysis on COVID-19 $R_0$ based on 23 studies conducted within China found the pooled estimate of $R_0$ to be 3.32 (Alimohamadi, Taghdir, & Sepandi, 2020). This essentially means that one COVID-19 patient in India infects lesser number of individuals in susceptible population in comparison to the global scenario. For a more profound assessment of this very significant parameter, we further determined the $R_0$ values for the months of May and June (01-20 June) 2020 for India. While $R_0$ value for April 2020 was 1.72, it declined to 1.40 and 1.29 for May and June (01-20 June) 2020, respectively. The evidence essentially suggests that not only the basic reproduction number of India is lower than the global estimates, it is declining and inching closer to one with each passing month. This could be attributed to two reasons. First, the robust policies implemented by the central and state governments of India for prevention and contact tracing. Some of the measures are - a strict lockdown imposed by Government of India during 25 March to 31 May 2020; large scale awareness programme to educate citizens; and, encouraging citizens to download "*Arogya Setu*", a dedicated app with proximity information on COVID-19. Second, extant research suggests possible greater immunity of Indians due to the various vaccination programs mandatorily implemented in India (Banik et al., 2020). Even as $R_0$ is still much larger than the value of 1 required for the epidemic to show signs of stagnation (Liu et al., 2020), yet it is an encouraging sign given the large population and high population density of India.

Another interesting finding of the study is pertaining to the lower case-fatality rate $\sigma$ of India. The value of $\sigma$ is determined using equation (2) and is based on $R_1$, the fraction of infected individuals who died. For India, $R_1$ is 3.22% as on 20[th] June 2020. Whereas, Onder, Rezza, and Brusferro (2020) suggest that the rate was 7.2% for Italy as on 17 March 2020, the fraction has increased to 14.5% as on 20 June 2020 (Worldometer, 2020). Similarly, for USA and Brazil, two of the most affected countries, the value of $R_1$ as on 20$^{th}$ June 2020 stands at 5.3% and 4.7%, respectively (Worldometer, 2020). India faces some unique demographic challenges in terms of a very large population, high population density, and scarcity of healthcare resources. In view of these challenges, it is very difficult on India's part to drastically contain the spread of COVID-19. More so when almost 80% of COVID-19 infected individuals in India are asymptomatic and may never have been known but for contact tracing (Times of India, 2020, May 28). Therefore, India should primarily invest its limited healthcare resources to limit $R_1$, the fraction of infected individuals who die.

The overall picture of COVID-19 pandemic presented by this study is that of an epidemic which may continue to impact us all the way up to the first quarter of 2021. As WHO has rightly pointed out, the worst of COVID-19 is yet to come since the pandemic is now in acceleration mode. Our study also reflects this observation (Figures 1 and 3). The pandemic is likely to reach its peak during October 2020 and may start showing signs of decline only in January 2021 in most parts of India (Table III). While the findings are critical, the study has two limitations which should not be overlooked. First, there is possibility of potential bias since different states have different testing capacities and reporting jurisdictions which cannot be accounted for in the model. Second, the model assumes that future would be a mirror image of the past. As data scientists we have extrapolated the past with a robust theoretical framework. Given the novelty of the virus, we do not know how some of the important epidemiological parameters like incubation period, mean serial interval, generation time would pan out in future. The study stands out as the only study which has projected COVID-19 confirmed cases for India with data as recent as 20$^{th}$ June 2020. It would help the policy makers, public health authorities, and industry leaders to prepare themselves better in the coming months in terms of leveraging healthcare and economic resources.

Times of India. Almost 80% of COVID cases in India asymptomatic: Union health minister. https://timesofindia.indiatimes.com/life-style/health-fitness/health-news/almost-80-of-covid-cases-in-india-asymptomatic-union-health-minister/articleshow/76069643.cms; 2020, May 28 [accessed 21 June 2029].

Toda AA. Susceptible-infected-recovered (SIR) dynamics of COVID-19 and economic impact, arXiv preprint arXiv:2003.11221; 2020 [accessed 19 June 2020].

Van Rossum G, Drake FL. PYTHON 2.6 Reference Manual; 2009.

Wallinga J, Lipsitch M. How generation intervals shape the relationship between growth rates and reproductive numbers. Proceedings of the Royal Society B: Biological Sciences; 2007, 274, 599-604.

WHO. WHO Coronavirus Disease (COVID-19) Dashboard. https://covid19.who.int/; 2020 [accessed 21 June 2020].

Worldometers. Coronavirus Cases. https://www.worldometers.info/coronavirus/country/india/; 2020 [accessed 21 June 2020].

Zhao S, Lin Q, Ran J, Musa SS, Yang G, Wang W, Lou Y, Gao D, Yang L, He D, Wang MH. Preliminary estimation of the basic reproduction number of novel coronavirus (2019-nCoV) in China, from 2019 to 2020: A data-driven analysis in the early phase of the outbreak. International Journal of Infectious Diseases; 2020, 92, 214-217.

Zhuang Z, Zhao S, Lin Q, Cao P, Lou Y, Yang L, Yang S, He D, Xiao L. Preliminary estimating the reproduction number of the coronavirus disease (COVID-19) outbreak in Republic of Korea and Italy by 5 March 2020. International Journal of Infectious Diseases; 2020, 95, 308-310.
20